\pdfoutput=1
\documentclass[11pt]{article}

\usepackage[utf8]{inputenc}
\usepackage[T1]{fontenc}

\usepackage{arxiv}

\usepackage{amsmath}
\usepackage{amssymb}
\usepackage{graphicx}
\usepackage{float}
\usepackage{booktabs}
\usepackage{listings}
\usepackage{subcaption}
\usepackage{enumitem}
\usepackage{natbib}

\usepackage{hyperref}
\usepackage{url}
\hypersetup{
  colorlinks=true,
  linkcolor=arxivaccent,
  citecolor=arxivaccent,
  urlcolor=arxivaccent,
  filecolor=arxivaccent,
  breaklinks=true,
}

\definecolor{codebg}{rgb}{0.95,0.95,0.97}
\definecolor{codegreen}{rgb}{0.0,0.5,0.0}
\definecolor{codegray}{rgb}{0.5,0.5,0.5}
\definecolor{codepurple}{rgb}{0.58,0,0.82}
\definecolor{codeblue}{rgb}{0.0,0.0,0.7}
\lstdefinestyle{codestyle}{
    backgroundcolor=\color{codebg},
    basicstyle=\ttfamily\small,
    breakatwhitespace=false,
    breaklines=true,
    captionpos=b,
    commentstyle=\color{codegreen},
    keywordstyle=\color{codeblue}\bfseries,
    numberstyle=\tiny\color{codegray},
    stringstyle=\color{codepurple},
    language=C,
    showstringspaces=false,
    numbers=left,
    numbersep=5pt,
    frame=single,
    rulecolor=\color{codegray},
    tabsize=4,
}
\title{Fine-Grained Computation Offload for\\Off-the-Shelf Servers in Tens of Lines}

\author{
  Bojie Li \\
  Pine AI
}

\date{}
\runningtitle{Fine-Grained Computation Offload in Tens of Lines}

\begin{document}
\maketitle

\begin{abstract}
Hardware accelerators now sit on the critical path of online serving---GPUs, FPGAs, and
increasingly \emph{remote} services such as hardware security modules, post-quantum KEMs, and
inference servers. For \emph{fine-grained} offloads (microseconds to a few milliseconds) the
classic responses to the resulting stall both fail: a context switch costs as much as the offload,
and a busy-wait burns the core. Overlapping the offload with other requests is the fix, and prior
systems obtain it by \emph{adding} concurrency: an async-framework rewrite, a new runtime or
dataplane OS, or a hand-tuned point integration.

We observe that the concurrency already exists: serving concurrent requests \emph{is} suspending
and resuming them, so every server ships the machinery overlap needs. Overlap is then a
\emph{routing} problem, not a rewrite problem: submit the offload to an executor, suspend the
request with the server's own deferred-response primitive, resume it on completion. Across ten
off-the-shelf servers spanning every production concurrency model, this recipe takes \textbf{22--138
lines added, at most one modified}, and recovers \textbf{$1.2$--$5.4\times$} on real hardware; the
server's concurrency model and the offload's weight predict both numbers in advance, and the win
is bounded by device throughput and the server's own overlap capacity. At the limit, an
\texttt{LD\_PRELOAD} fiber runtime injects the reroute into an unmodified thread-per-connection
binary ($17.3\times$) within a characterized envelope. Rerouting suspends
run-to-completion atomicity; a measured taxonomy confines the hazard to unlocked shared
aggregates, and a transparent page-protection detector guards exactly those, validated on stock
Redis.
\end{abstract}

\begin{center}
\small
Code: \url{https://github.com/19PINE-AI/transparent-offload} \\[2pt]
Website: \url{https://01.me/research/transparent-offload}
\end{center}
\vspace{-0.5em}

\begin{figure}[H]
  \centering
  \includegraphics[width=\linewidth]{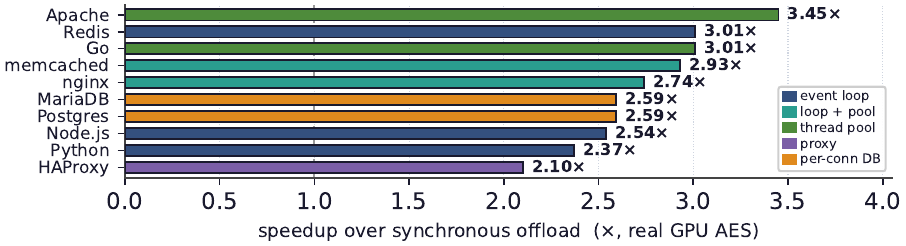}
  \caption{The headline result. Rerouting the offload through each server's own concurrency
    recovers $2.1$--$3.5\times$ across ten servers (real GPU AES, $1$\,MiB; up to $5.4\times$ at
    $8$\,MiB; the databases pipeline the op intra-query); zero-edit rerouting from outside the
    binary reaches $17.3\times$ (\S\ref{sec:transparent}). Color marks the concurrency model.}
  \label{fig:headline}
\end{figure}

% =====================================================================
%  body.tex  —  Fine-Grained Computation Offload for Off-the-Shelf Servers
%               in Tens of Lines
% =====================================================================

\section{Introduction}
\label{sec:intro}

Hardware accelerators are now a standard part of online serving. GPUs, TPUs, and FPGAs accelerate
inference, image processing, encryption, and compression, and a growing share of ``acceleration''
is in fact a \emph{remote} call: to a hardware security module, a post-quantum key-encapsulation
service~\citep{mlkem}, or a co-located inference server~\citep{clipper}. Across these cases the
application has the same shape: receive a request, pre-process, invoke an intensive routine,
post-process, respond. On an accelerator that routine becomes an \emph{offload}: a submission to
the device and a wait for its completion, lasting microseconds to a few milliseconds. This paper
is about a deceptively simple question: \emph{what should the CPU do while it waits?}

\begin{figure}[t]
  \centering
  \includegraphics[width=\linewidth]{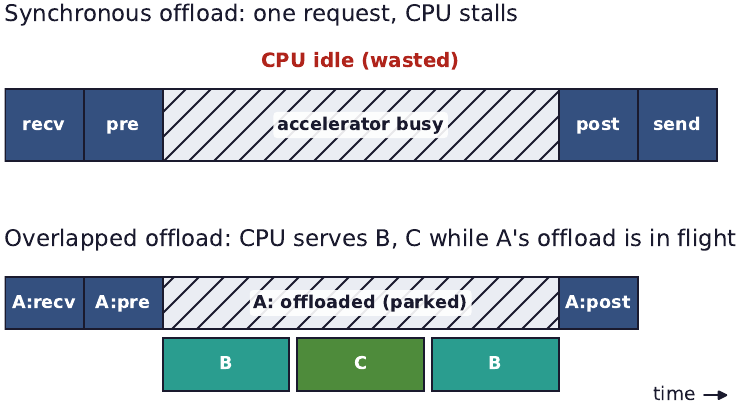}
  \caption{The problem. A serving request is \emph{receive $\to$ pre-process $\to$ offload $\to$
    post-process $\to$ send}. Top: with a synchronous offload, the CPU stalls while the accelerator
    works. Bottom: \emph{overlap} fills the gap with other requests' processing.}
  \label{fig:pipeline}
\end{figure}

The textbook answers fail exactly in this regime (Figure~\ref{fig:pipeline}). \emph{Blocking} lets
the operating system run another thread, but a context switch costs several microseconds,
comparable to the offload itself; this is the ``killer microsecond''~\citep{microsecond}: the
switch wastes CPU \emph{and} adds a wakeup delay to the request's tail. \emph{Busy-waiting} hides
the wakeup but burns a core doing nothing. The only principled answer is to \emph{overlap} the
offload with the processing of other requests.

Overlap requires concurrency at the offload point, and the systems community has treated that
concurrency as something to be \emph{added}: write the application against an asynchronous
framework from the start~\citep{libevent,seastar}, adopt a new runtime or dataplane operating
system~\citep{ix,shenango,caladan}, or hand-integrate one offload into one server~\citep{sslshader,qat}.
All of these assume the server lacks what overlap needs.

Our starting observation is that \textbf{it already has it}. Serving concurrent connections
\emph{is} suspending and resuming requests: an event loop parks a connection whenever it waits for
a socket; a worker pool parks whole requests in its queue; a goroutine scheduler parks tasks at
every blocking call; a database parks each connection in its own backend. Every modern server
therefore ships, in production quality, exactly the machinery that overlap requires. Hiding a
fine-grained offload is not a rewrite problem but a \textbf{routing problem}: at the offload call
site, \emph{submit} the work to a background executor, \emph{suspend} the request using the
server's own deferred-response primitive, and \emph{resume} it to reply on completion.

We validate this recipe on \textbf{ten} off-the-shelf servers spanning every concurrency model in
production use: Redis, Node.js, Python/asyncio, nginx, memcached, Apache, Go, HAProxy, PostgreSQL,
and MariaDB. Across the ten, the change is \textbf{22--138 lines added and zero or one existing
line modified}---the core of the Redis integration is a dozen lines
(Listing~\ref{lst:redis})---and on real hardware it recovers $1.2$--$5.4\times$ across servers
and offload weights (Figure~\ref{fig:headline}). The cost is small for a structural reason, not a lucky one:
the integration only bridges the offload to machinery that already exists.

That structure makes the outcome \emph{predictable}. Two properties of the deployment fix both the
speedup and the code cost in advance: the server's \emph{concurrency model} determines what a
synchronous offload costs and therefore what rerouting recovers, and how many lines the reroute
takes; the offload's \emph{weight relative to per-request CPU work} determines whether overlap
pays at all. The model locates every server we measured, including the ones where overlap does
\emph{not} help.

Two questions remain, and they complete the paper. First, \emph{can the change be zero?} When
there is no source access at all, the libc symbol boundary is itself a suspension layer: an
\texttt{LD\_PRELOAD} fiber runtime injects the rerouting from outside an unmodified binary and
reaches $17.3\times$ on a thread-per-connection server---within an envelope we characterize
precisely, because transparency extends exactly as far as the behavior that flows through the
interposed layer. Second, \emph{what does rerouting break?} Suspending a handler mid-request
forfeits the run-to-completion atomicity that event-driven servers silently rely on---a hazard we
measure, scope, and guard.

In summary, this paper makes three contributions:
\begin{itemize}[leftmargin=1.4em,itemsep=2pt,topsep=2pt]
  \item \textbf{An offload-rerouting method.} Overlapping a fine-grained offload requires rerouting it
    through concurrency the server already has, not adding concurrency. One three-step method,
    instantiated on ten off-the-shelf servers with 22--138 lines added and at most one
    existing line modified,
    recovers $1.2$--$5.4\times$ on real hardware across offload weights (\S\ref{sec:recipe},
    \S\ref{sec:eval}).
  \item \textbf{A predictive model of speedup and code cost.} Concurrency model and offload weight
    predict the win and the cost in advance (\S\ref{sec:model}); at the zero-edit limit, an
    interposition-envelope principle and a syscall-profile classifier predict where transparent
    rerouting is possible at all (\S\ref{sec:transparent}).
  \item \textbf{A correctness taxonomy and a transparent guard.} Rerouting breaks run-to-completion
    atomicity; a measured taxonomy confines the hazard to unlocked shared aggregates, and a
    transparent conflict detector protects exactly those with no application changes
    (\S\ref{sec:correctness}).
\end{itemize}

% =====================================================================
\section{The Fine-Grained Regime and a Predictive Model}
\label{sec:model}

\begin{figure}[t]
  \centering
  \includegraphics[width=\linewidth]{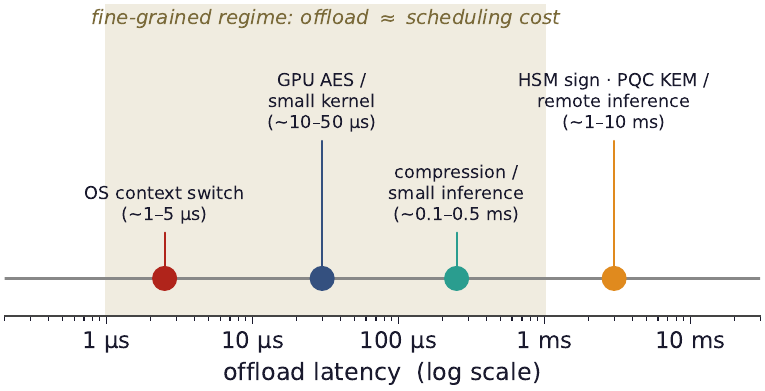}
  \caption{The fine-grained regime. Accelerator offloads span microseconds to milliseconds. In the
    shaded band, the offload latency is comparable to an OS context switch, so blocking pays a switch
    and a wakeup that rival the work itself, and busy-waiting wastes a core.}
  \label{fig:landscape}
\end{figure}

\paragraph{The fine-grained regime.}
An accelerator turns a CPU-bound routine into a device operation: the CPU prepares an input
buffer, submits it (a kernel launch, a DMA, or a network send), and later collects the result.
These round-trips span microseconds to milliseconds (Figure~\ref{fig:landscape}), but all are
\emph{fine-grained} relative to operating-system scheduling: the offload finishes on the timescale
of a context switch, so the dilemma of \S\ref{sec:intro} holds across the whole band.

\paragraph{A taxonomy of concurrency models.}
The thesis says to reroute the offload through the server's own concurrency, so the first question
about any server is where that concurrency lives. Five models cover production practice. A
\emph{single-event-loop} server (Redis, Node.js, a Python \texttt{asyncio} service) multiplexes all
connections on one thread; its suspension primitive is the deferred reply. An
\emph{event-loop-with-pool} server (nginx, memcached) runs a few such loops plus a worker pool. A
\emph{thread- or goroutine-pool} server (Apache, Go) parks whole requests in pooled workers or
runtime-scheduled tasks. A \emph{proxy} (HAProxy) can route requests through a native offload
engine to an external agent. A \emph{process- or thread-per-connection} server (PostgreSQL,
MariaDB) gives each connection its own backend, which the OS suspends and resumes wholesale.

\begin{figure}[t]
  \centering
  \includegraphics[width=\linewidth]{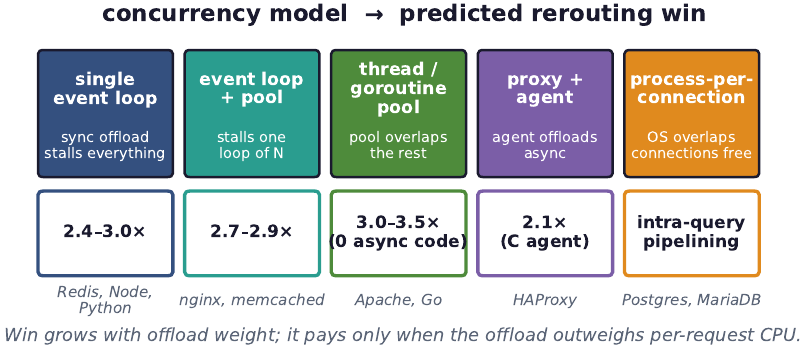}
  \caption{Prediction 1: the concurrency model determines what a synchronous offload costs and
    therefore how much rerouting recovers, from \emph{dramatic} (a blocked single loop) to
    \emph{automatic} (a pool that overlaps for free) to \emph{intra-query} (databases that already
    overlap across connections). Example servers are shown under each regime.}
  \label{fig:regimes}
\end{figure}

\paragraph{Prediction 1: concurrency model determines speedup and code cost.}
The same taxonomy predicts what a \emph{synchronous} offload costs and therefore what rerouting
recovers (Figure~\ref{fig:regimes}). The extreme case is the single event loop: one thread handles
every connection, so a synchronous offload stalls \emph{all} of them and throughput collapses to
one request per offload latency while the hardware idles---a pathology of \emph{structure}, not
capacity---and the few-line reroute is dramatic there. In an event loop with a pool, a synchronous
offload stalls only one loop of several. A thread or goroutine pool overlaps offloads
\emph{automatically}: the reroute needs no asynchronous code at all, because parking the worker
\emph{is} the routing. For a proxy, the reroute is configuration plus an external agent. And a
per-connection database already overlaps \emph{across} connections for free (the OS runs the
backends concurrently), leaving \emph{intra-query} pipelining of a serial offload loop as the only
win.

The \emph{code cost} follows the same axis: where the server exposes a plugin API or a
deferred-response primitive the reroute is purely additive; where it exposes none, the
developer must add the suspend/resume transition to the request state machine by hand, which is
where the line count comes from (\S\ref{sec:recipe}).

\paragraph{Prediction 2: offload weight determines whether overlap pays.}
The second property is independent of the server. Overlap reclaims the CPU time the offload would
have wasted, so if per-request CPU work already rivals the offload there is little to reclaim; the
win grows with the offload's weight until it hits one of two ceilings: the
\emph{accelerator's throughput} (overlap fills the device but cannot exceed it) or the
\emph{server's own overlap capacity} (the concurrency its machinery can keep in flight). Real
speedups live between these bounds. Section~\ref{sec:eval} confirms both predictions, including a
block-size sweep that traces the weight dependence on real hardware (Figure~\ref{fig:weight}) and
the servers where overlap correctly buys nothing.

Together the two properties form a map: given a server's concurrency model and an offload's
latency, an operator can predict before writing any code whether to bother, how large the win will
be, and roughly how many lines it will take.

% =====================================================================
\section{Offload Rerouting on Ten Servers}
\label{sec:recipe}

\begin{figure}[t]
  \centering
  \includegraphics[width=\linewidth]{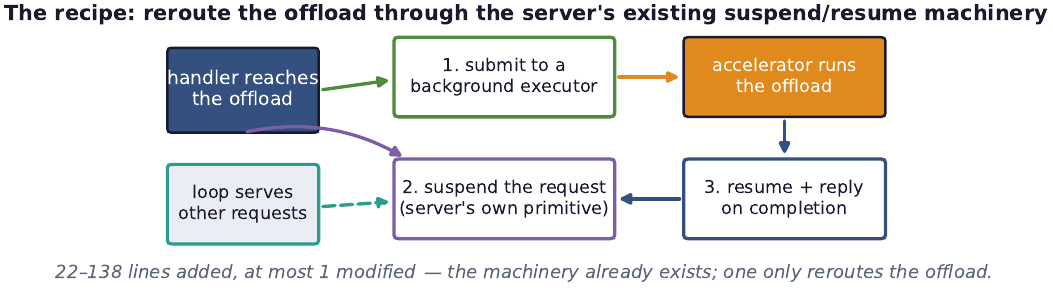}
  \caption{The rerouting method. While a request is suspended at its offload, the server's loop
    serves other requests. The edit is small because the suspend/resume machinery already
    exists.}
  \label{fig:recipe}
\end{figure}

The recipe is the thesis made concrete, and it is uniform across servers
(Figure~\ref{fig:recipe}):

\begin{enumerate}[leftmargin=1.5em,itemsep=1pt,topsep=2pt]
  \item at the offload call site, \emph{submit} the work to a background executor instead of waiting;
  \item \emph{suspend} the current request using the server's own deferred-response primitive;
  \item \emph{resume} it and send the reply when the offload completes.
\end{enumerate}

Listing~\ref{lst:redis} shows the recipe in Redis: the entire integration is an 83-line loadable
module with zero edits to Redis itself, and its overlap core is the dozen lines of the listing. The
synchronous variant of the same command simply calls \texttt{accel\_encrypt} on the event-loop
thread; the diff between the two \emph{is} the reroute.

\begin{lstlisting}[caption={The reroute in Redis (condensed from the 83-line module
\texttt{accel\_module.c}). \texttt{BlockClient} is Redis's own deferred-response primitive; the
worker pool is the background executor. The command suspends before submitting, so the resume
cannot race the suspension.},label={lst:redis},float=t]
/* accel.async: submit, suspend, resume on completion */
int cmd_async(RedisModuleCtx *ctx, ...) {
    RedisModuleBlockedClient *bc =
        RedisModule_BlockClient(ctx, reply_cb,
                    timeout_cb, NULL, 0); /* 2. suspend */
    enqueue(bc);                          /* 1. submit  */
    return REDISMODULE_OK;  /* loop serves other clients */
}
void *worker(void *arg) {                 /* pool thread */
    for (;;) { job_t j = dequeue();
        accel_encrypt(j.buf, j.len);      /* the offload */
        RedisModule_UnblockClient(j.bc, NULL);
    }                                     /* 3. resume   */
}
\end{lstlisting}

\begin{table}[t]
  \centering
  \small
  \begin{tabular}{llrrr}
    \toprule
    server & integration point & +lines & mod & speedup \\
    \midrule
    Redis 6.0        & loadable module (\texttt{BlockClient})     & 83  & 0 & $3.01\times$ \\
    Node.js 22       & N-API add-on (libuv queue)                 & 34  & 0 & $2.54\times$ \\
    Python 3.10      & \texttt{run\_in\_executor}                 & 22  & 0 & $2.37\times$ \\
    nginx 1.18       & add-on module (thread pool + aio)          & 112 & 0 & $2.74\times$ \\
    memcached 1.6    & state-machine patch                        & 70  & 1 & $2.93\times$ \\
    Apache 2.4       & module (\texttt{apxs}), pooled workers     & 27  & 0 & $3.45\times$ \\
    Go 1.18          & plain blocking \texttt{cgo} call           & 28  & 0 & $3.01\times$ \\
    HAProxy 2.4      & SPOE + standalone C agent                  & 138 & 0 & $2.10\times$ \\
    PostgreSQL 14    & C extension (intra-query)                  & 42  & 0 & $2.59\times^\dagger$ \\
    MariaDB 10.6     & UDF (intra-query)                          & 34  & 0 & $2.59\times^\dagger$ \\
    \bottomrule
  \end{tabular}
  \caption{Offload rerouting on ten off-the-shelf servers (grouped by concurrency model: single event
    loop, event loop + pool, thread/goroutine pool, proxy, per-connection database). ``+lines'' is
    integration code added; ``mod'' is existing lines modified. Speedups are over the synchronous
    offload with a real GPU ($1$\,MiB AES; $^\dagger$the databases pipeline the op intra-query:
    serial vs.\ pipelined offloads within one query). Seven are stock server binaries; for Node.js, Python, and Go the server is an
    idiomatic handler on the stock runtime. Measurement setup: \S\ref{sec:eval}.}
  \label{tab:servers}
\end{table}

Table~\ref{tab:servers} instantiates the recipe across the landscape, and the integration point
follows the concurrency model exactly as \S\ref{sec:model} predicts. Servers with a \emph{plugin
or extension API} take the reroute with zero core edits: a Redis~\citep{redis} module, an
nginx~\citep{nginx} thread-pool add-on, an Apache~\citep{apache} content handler,
PostgreSQL~\citep{postgresql} and MariaDB~\citep{mariadb} user-defined functions, and a
Node.js~\citep{nodejs} N-API add-on over libuv~\citep{libuv}. Servers backed by a \emph{language
runtime}, whatever their concurrency model, need almost no new code: the goroutine scheduler
overlaps a Go~\citep{golang} handler's plain blocking \texttt{cgo} offload with \emph{zero}
asynchronous code, and a Python \texttt{asyncio}~\citep{cpython} service reroutes through
\texttt{run\_in\_executor} in one line of handler code. A \emph{proxy} reroutes in
configuration: HAProxy~\citep{haproxy} streams each request over its native Stream Processing
Offload Engine (SPOE)~\citep{spoe} to a standalone offload agent (the proxy itself changes only configuration;
the 138-line C agent carries the offload).

The lone case that touches existing code is
memcached~\citep{memcached}, which exposes no deferred-response primitive, so we add the
suspend/resume transition to its connection state machine by hand (70 lines, one modified). It is
the exception that proves the structural claim: \textbf{the cost of the reroute is the distance to
the server's nearest suspend/resume primitive}.

\begin{figure}[t]
  \centering
  \includegraphics[width=\linewidth]{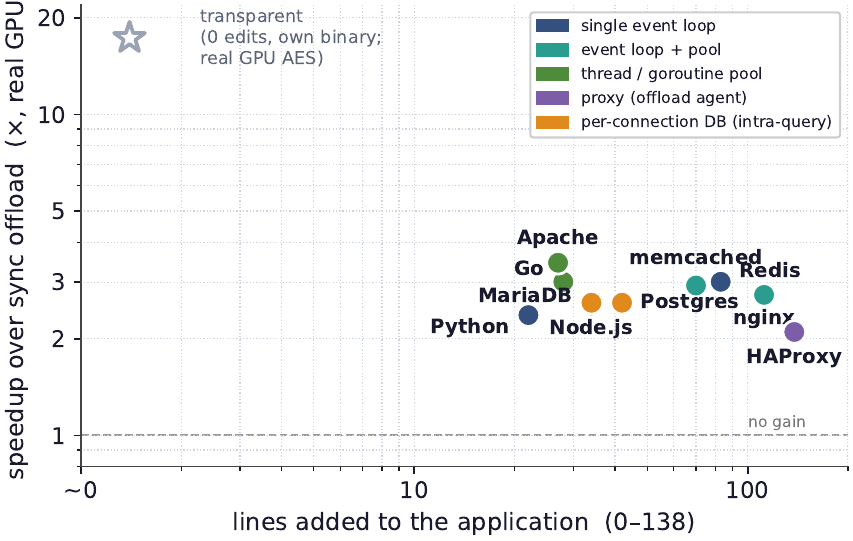}
  \caption{The study in one plot: lines of integration code vs.\ speedup from rerouting a real-GPU
    AES offload on an idle device ($1$\,MiB blocks; the databases pipeline the op intra-query).
    The few-line reroute clusters at $2$--$3.5\times$; the zero-edit transparent runtime
    reaches $17.3\times$ in its niche
    (\S\ref{sec:transparent}). Color encodes the concurrency model.}
  \label{fig:spectrum}
\end{figure}

We report \emph{lines added} separately from \emph{existing lines modified} because they carry
different maintenance costs: added lines live in a module or extension and survive server
upgrades; modified lines are the invasive part. Figure~\ref{fig:spectrum} plots the whole study on one
canvas: code cost against measured win, with the zero-edit limit of \S\ref{sec:transparent} at the
origin.

% =====================================================================
\section{The Zero-Edit Limit: Rerouting by Interposition}
\label{sec:transparent}

\begin{figure}[t]
  \centering
  \includegraphics[width=0.92\linewidth]{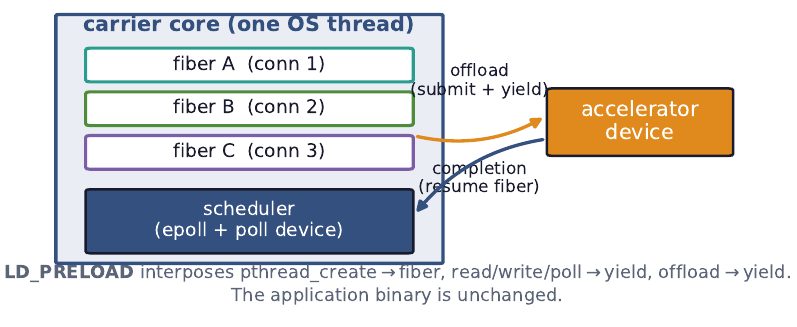}
  \caption{The zero-edit limit. An \texttt{LD\_PRELOAD} library turns each connection thread into
    a fiber on one carrier thread; a fiber yields at its offload (and at socket I/O) and the
    scheduler runs another until the offload completes. The binary is unchanged. On a synchronous
    thread-per-connection handler with a real GPU this overlaps 64 connections' offloads for a
    $17.3\times$ gain over a busy-wait synchronous baseline.}
  \label{fig:runtime}
\end{figure}

Can the reroute cost \emph{zero} lines? When there is no source access at all---a stock binary, a
vendor blob---the recipe cannot be applied from inside. But there is one suspension layer every
dynamically linked binary passes through: the libc symbol boundary. Our transparent runtime is a
shared library loaded by \texttt{LD\_PRELOAD} that interposes the standard threading and I/O
symbols. When the application creates a connection-handling thread, the runtime instead creates a
\emph{fiber}---a user-level thread with a register-only context switch of tens of
nanoseconds~\citep{fastwake}---and multiplexes all fibers on one \emph{carrier} OS thread
(occupying a single core). Whenever a
fiber would block, on socket I/O or on the offload, the runtime switches to another runnable
fiber; a scheduler polls I/O readiness and offload completions and resumes the corresponding
fiber, saving and restoring per-fiber libc state such as \texttt{errno} across switches. The
handler keeps its plain synchronous shape and never learns that its ``thread'' is a fiber: the
recipe's submit, suspend, and resume are injected from outside the binary
(Figure~\ref{fig:runtime}).

On its home ground this works well: on a synchronous thread-per-connection handler with a real GPU
performing AES, the runtime overlaps 64 connections' offloads on one carrier thread and reaches
$17.3\times$ the throughput of the same binary busy-waiting on each offload, with results verified
bit-for-bit ($11.9\times$ over a baseline that blocks in the driver and lets the OS overlap the 64
threads\footnote{The blocking-baseline comparison was measured while a co-tenant shared the GPU
(both sides equally affected); the busy-wait comparison ran on an otherwise idle GPU.}); a
separately built DNN-inference server shows $11.8\times$.

Making stock binaries run under the
runtime took a catalog of interposition engineering---most notably a glibc condition-variable
symbol-versioning hazard that silently corrupts some binaries and that any threading-interposing
tool must fix---which Appendix~\ref{app:obstacles} records for practitioners.

\begin{figure}[t]
  \centering
  \includegraphics[width=\linewidth]{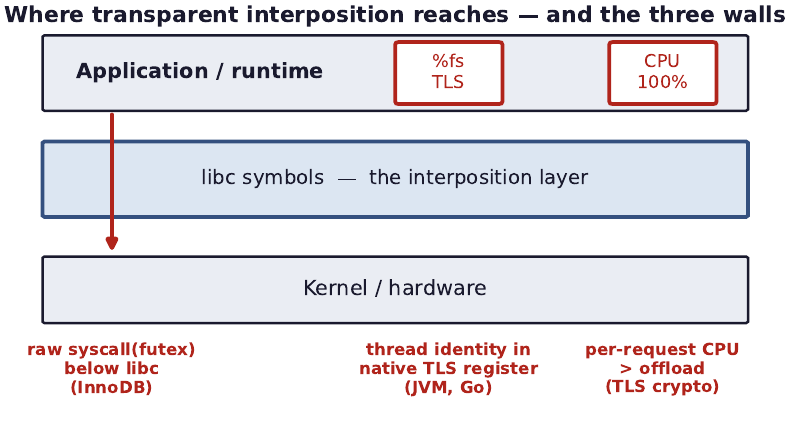}
  \caption{The interposition envelope. A preloaded library virtualizes the libc symbol layer, and
    reaches exactly the behavior that flows through it. Behavior escapes in three ways: below the
    layer (raw \texttt{futex} syscalls, InnoDB), beside it (thread identity in a native TLS
    register, the JVM and Go), and behind it (per-request CPU that already exceeds the offload).}
  \label{fig:walls}
\end{figure}

\paragraph{The interposition envelope.}
The more useful contribution is the limit. A preloaded library virtualizes exactly one layer, the
libc symbol boundary, so \textbf{its reach is the completeness of that layer}: behavior that
escapes the layer is out of reach, and there are exactly three ways to escape it
(Figure~\ref{fig:walls}).
\begin{itemize}[leftmargin=1.4em,itemsep=2pt,topsep=2pt]
  \item \emph{Below it.} A storage engine such as InnoDB synchronizes with raw
    \texttt{futex} system calls~\citep{futexes} issued directly, bypassing \texttt{pthread}; a fiber
    that blocks in a raw futex stalls the whole carrier, and under contention the server
    deadlocks---we confirmed this with a live backtrace of the frozen carrier inside InnoDB.
  \item \emph{Beside it.} A managed runtime such as the JVM keeps ``the current thread'' in a native
    thread-local slot read inline from a CPU register, which symbol interposition cannot virtualize
    per fiber; the moment two fiberized JVM requests interleave, the runtime reads the wrong thread
    object and segfaults (the same mechanism applies to Go and .NET).
  \item \emph{Behind it.} Overlap needs idle
    CPU to reclaim, and a thread-per-connection TLS terminator that spends real CPU on per-request
    crypto has none---the OS already overlaps its offloads across threads, while the single carrier
    serializes the crypto and pays an interposition tax on every yield, ending up $2$--$3\times$
    \emph{slower} than native.
\end{itemize}
These are properties of where an application places its behavior, not
gaps in engineering; no interposition removes them.

\begin{figure}[t]
  \centering
  \includegraphics[width=0.86\linewidth]{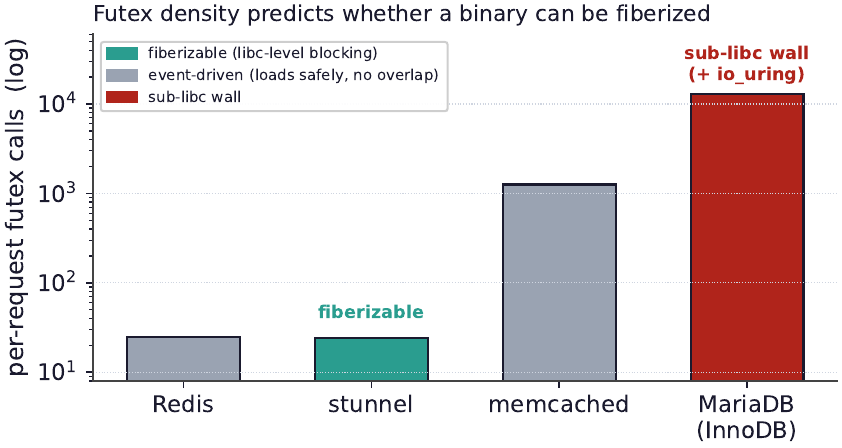}
  \caption{Predicting the envelope. Per-request \texttt{futex} calls separate
    thread-per-connection servers that block at the libc layer (fiberizable: stunnel) from engines
    that synchronize below it (InnoDB, $\sim$$13{,}000$, some $500\times$ higher, plus
    \texttt{io\_uring}). Event-driven servers (gray) have no per-connection thread to fiberize;
    their \texttt{epoll} profile flags them ``loads safely, no overlap.''}
  \label{fig:classifier}
\end{figure}

\paragraph{A syscall-profile classifier.}
The envelope is predictable from the outside: we profile the per-request blocking syscalls a
server makes under load (Figure~\ref{fig:classifier}). Event-driven servers show an
\texttt{epoll}-dominated profile and have no per-connection thread to fiberize: the runtime loads
safely but never engages. Thread-per-connection servers that block at the libc layer show
near-zero per-request \texttt{futex} traffic and are fiberizable. Engines that synchronize below
libc are unmistakable: InnoDB issues $\sim$$13{,}000$ futex calls per request, some $500\times$ a
fiberizable server's, alongside asynchronous kernel I/O (\texttt{io\_uring}~\citep{iouring}). A cheap
\texttt{strace} therefore tells an operator in advance whether zero-edit rerouting can work.

The verdict frames the division of labor. Zero-edit rerouting serves a real but narrow niche:
thread-per-connection servers with light per-request CPU and libc-level blocking, reached with no
source access. By construction it cannot help event-driven servers---the class where a synchronous
offload is most catastrophic. For everyone else, the few-line recipe of \S\ref{sec:recipe} is the
answer.

% =====================================================================
\section{Correctness under Rerouting}
\label{sec:correctness}

\begin{figure}[t]
  \centering
  \includegraphics[width=\linewidth]{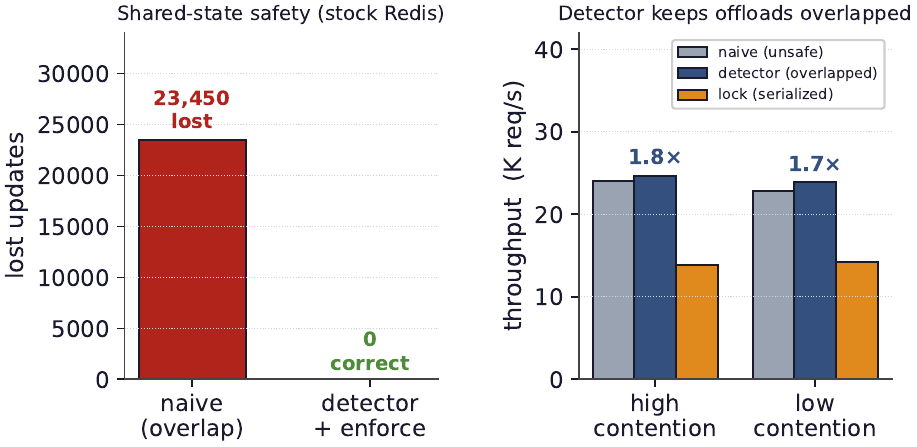}
  \caption{Guarding the one hazardous pattern, measured on a stock Redis server with a real GPU
    offload. Left: unprotected overlap of a shared read-modify-write loses $23{,}450$ updates. The
    page-protection detector flags $26{,}185$ conflicts and, under enforcement, reduces lost updates
    to zero. Right: a coarse lock serializes all offloads ($13.8$K\,req/s), while the
    detector keeps non-conflicting offloads overlapped ($24.7$K\,req/s), $1.8\times$ faster
    ($1.7\times$ at low contention) and within noise of unprotected overlap ($24.1$K\,req/s).}
  \label{fig:correctness}
\end{figure}

Rerouting has one cost that is easy to miss. A single-threaded server runs each handler to
completion, so handlers observe shared state atomically---an invariant the code \emph{silently}
relies on. Suspending a handler at its
offload breaks that invariant: if post-processing does a read-modify-write on shared state---a
counter, a cache entry, a key in a store---two overlapped requests can interleave their sequences
and lose updates. On a \emph{stock Redis server} whose module command reads a key, runs a real GPU
offload, and writes the incremented value back, unprotected overlap silently loses tens of
thousands of updates (Figure~\ref{fig:correctness}, left). The more latency overlap hides, the
wider the race it opens.

\paragraph{A measured taxonomy of shared-state patterns.}
To scope the problem we enumerated the shared-state patterns that real offload-adjacent servers
keep, and measured each under overlapped execution in race-instrumented harnesses spanning crypto
(OpenSSL) and compression (zlib). Three patterns are unconditionally safe: \emph{read-only} shared
state (model weights, dictionaries, lookup tables), \emph{per-connection} state (which one handler
per connection serializes for free), and \emph{lock-protected} state (locks serialize regardless
of scheduling). All measured zero conflicts. The only hazardous pattern is the fourth:
\emph{unlocked shared mutable aggregates}---counters, batch queues, caches---where unlocked
read-modify-writes lose updates essentially every time they collide. And that pattern is exactly
the state a single-threaded server keeps unlocked \emph{because} it trusts run-to-completion
atomicity---the very atomicity rerouting removes. The dividing line is the state pattern, not the
application domain: bulk crypto, compression, hashing, and stateless inference land safe because
the heavy routine is pure and their shared state is read-only or already locked.

\paragraph{A transparent conflict detector.}
The safe patterns need nothing. For the residual---unlocked shared aggregates---a
transparent \emph{conflict detector} restores correctness with no application change: it
write-protects the shared-state pages, snapshots a version clock when a handler parks at its
offload, and treats a write fault on a page modified during the park as a conflict; under
enforcement it serializes only the conflicting handlers. On the stock-Redis workload it drives
lost updates to \emph{zero} while running $1.8\times$ faster than a coarse lock, which serializes
the very offloads it protects; its overhead versus unprotected overlap is within measurement noise
($24.7$K vs.\ $24.1$K\,req/s) (Figure~\ref{fig:correctness}, right). The detector is a property of
the overlap mechanism, not of any one integration, so it applies at every point of the spectrum,
including the zero-edit limit.

% =====================================================================
\section{Evaluation}
\label{sec:eval}

This section validates the two predictions of \S\ref{sec:model} on real hardware---Prediction~1
across the ten-server landscape, Prediction~2 by sweeping the offload's weight---and closes with
latency under load.

\begin{figure}[t]
  \centering
  \includegraphics[width=0.86\linewidth]{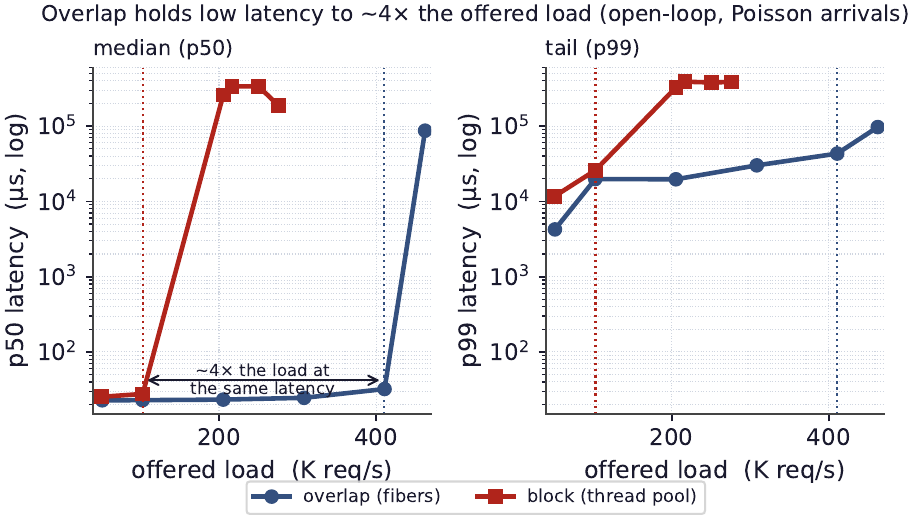}
  \caption{Latency under load (open-loop Poisson arrivals over a controlled $20$\,\textmu s
    \emph{emulated} offload, the one emulated experiment in the paper, used so offered load can be
    swept precisely; \texttt{runtime/openloop.csv}).
    Blocking on the offload drives both median (p50, left) and tail (p99, right) latency up at its
    saturation point near $103$K\,req/s, while overlapping holds low latency to roughly $4\times$
    that offered load (knee near $410$K\,req/s) before its own knee.}
  \label{fig:latency}
\end{figure}

\paragraph{Experimental setup.}
All cross-server offloads run on real hardware, with no emulated latencies; the one
exception is the open-loop latency study of Figure~\ref{fig:latency}, which uses a controlled
emulated offload and is labeled as such. Experiments run on a server with an NVIDIA RTX PRO 6000
(Blackwell) GPU. The GPU path performs AES~\citep{openssl} on its own CUDA stream, polled by
\texttt{cudaEventQuery}~\citep{cudastreams}, with the block size sweeping from $4$\,KiB
(launch-bound) to $8$\,MiB (bandwidth-bound). GPU AES is a stand-in for offloads that beat the
host CPU (a modern core's AES-NI rivals a GPU on this cipher); we use it because it gives a
cleanly tunable offload weight. For the latency-bound \emph{remote} class (HSM
signatures, post-quantum KEMs, remote inference) we stand up a real TCP signer doing a genuine
RSA-2048 signature per request. Servers are driven by standard load generators
(\texttt{redis-benchmark}, \texttt{ab}).

\paragraph{Validation of Prediction 1 across the server landscape.}
On a $1$\,MiB GPU AES offload (realistic bulk crypto, idle GPU) the reroute recovers the win the
model predicts in every regime, with no failed requests (Figure~\ref{fig:headline},
Table~\ref{tab:servers}): single event loops and event-loop-with-pool servers cluster around
$2.4$--$3\times$, thread and goroutine pools reach $3.0$--$3.5\times$ with \emph{no} asynchronous
code, and the per-connection databases recover $2.6\times$ through the one channel the model
leaves them, pipelining the offloads \emph{within} a query rather than across connections.
The cluster tops out at $2$--$3.5\times$ because this offload is bandwidth-bound: once
overlap saturates the device, its throughput is the ceiling. The proxy isolates where the win comes from:
HAProxy's gain rides entirely on the offload \emph{agent}, so a 21-line GIL-bound Python agent
gains nothing while a 138-line C \texttt{pthread}-pool agent recovers $2.1\times$ on the same
offload.

\paragraph{Validation of Prediction 2: offload weight and the two ceilings.}
Sweeping the GPU AES offload by block size on a single event loop traces the weight dependence
(Figure~\ref{fig:weight}; full table in Appendix~\ref{app:blocksize}): the same reroute gives
essentially nothing for a light block ($1.24\times$ at $4$\,KiB, $46$\,\textmu s, where the offload
is on the order of the server's own per-request work) and rises to $5.41\times$ at $8$\,MiB
($2.3$\,ms), approaching the GPU's bandwidth---the first ceiling, a property of the device. A
\emph{latency-bound} offload lifts that ceiling because its device is not saturated: with our real
RSA-2048 remote signer ($821$\,\textmu s round-trip) the same Python server reaches
$3.53\times$---but not the order of magnitude a deep queue could in principle give, because the
bottleneck moves to the second ceiling, the server's own overlap capacity (throughput and
round-trip latency imply the \texttt{asyncio}/GIL executor keeps only $\sim$$6$ requests truly in
flight). Real overlap is thus bounded at both ends; across our servers and offloads it lands at
$1.2$--$5.4\times$.

\begin{figure}[t]
  \centering
  \includegraphics[width=0.86\linewidth]{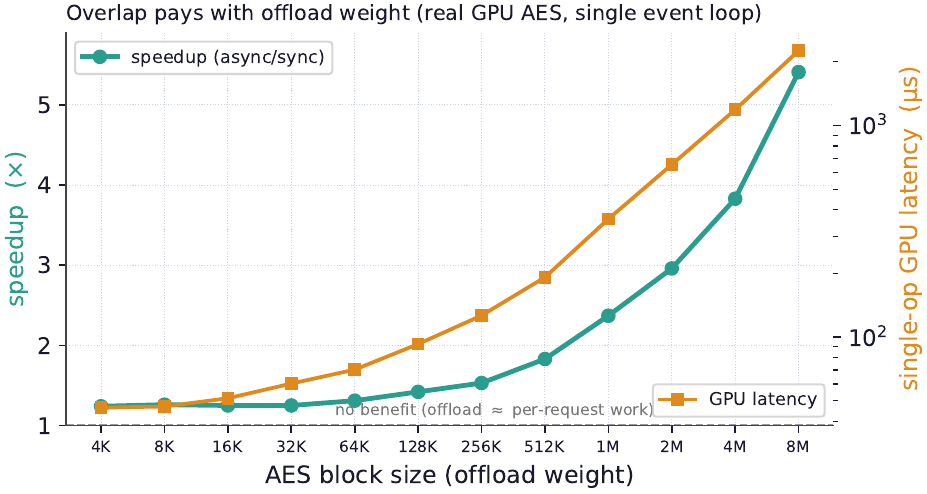}
  \caption{Prediction 2, measured (real GPU, idle): overlap pays only when the offload outweighs
    per-request CPU work. On a single-event-loop server, the same reroute gives $1.24\times$ for a
    light $4$\,KiB block ($46$\,\textmu s) and $5.41\times$ at $8$\,MiB ($2.3$\,ms), approaching
    the GPU's bandwidth. Right axis: measured single-op GPU latency.}
  \label{fig:weight}
\end{figure}

\paragraph{Latency under offered load.}
Overlap is a latency win too. By keeping the CPU busy during offloads, the overlapping path holds
both median and tail (p99) latency low up to roughly \emph{four times} the offered load that
blocking can sustain before its knee (Figure~\ref{fig:latency}).

% =====================================================================
\section{Discussion and Limitations}
\label{sec:discussion}

\paragraph{When rerouting does not pay.}
The model doubles as a stop sign. When per-request CPU already rivals the offload
(Prediction~2---the same condition that raises the transparent runtime's third wall), rerouting
buys complexity for nothing. And an operator whose accelerator is already saturated should buy
device bandwidth, not rewire the server: overlap recovers wasted utilization but never
manufactures throughput.

\paragraph{Scope of the conflict detector.}
The conflict detector is a targeted safety net for the one pattern rerouting endangers, not a
general transactional memory; shared state outside the monitored segments, or correctness
requirements beyond last-writer-wins, need stronger machinery.

% =====================================================================
\section{Related Work}
\label{sec:related}

\begin{figure}[t]
  \centering
  \includegraphics[width=0.86\linewidth]{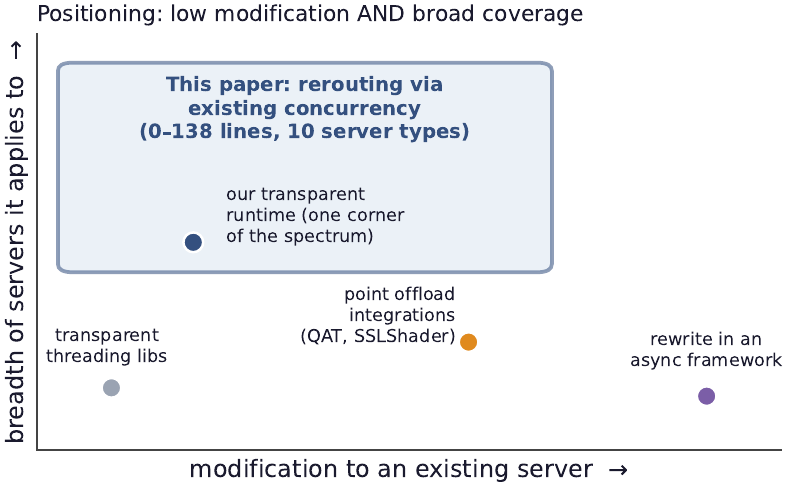}
  \caption{Positioning. Existing approaches are either low-effort but narrow (threading libraries,
    of which our own transparent runtime is one) or high-effort and still narrow (point
    integrations, async-framework rewrites). Rerouting through the server's own concurrency reaches the desirable
    region: little modification to existing servers, across many server types.}
  \label{fig:positioning}
\end{figure}

Figure~\ref{fig:positioning} places this work among prior approaches, which either \emph{add} the
concurrency---new frameworks, runtimes, and operating systems---or hand-integrate one offload into
one server; we instead reuse the concurrency existing servers already have, across the whole
landscape of server architectures, and predict the cost of doing so.

\paragraph{Threads, events, and coroutines.}
How to hide I/O latency cheaply is an old debate between an event-driven camp that structures the server
as a state machine over an event loop at the cost of ``stack ripping''~\citep{adya}
(Flash~\citep{flash}, SEDA~\citep{seda}) and a user-level-thread camp that keeps the synchronous
style and yields at blocking points (Capriccio~\citep{capriccio}, State
Threads~\citep{statethreads}, \texttt{libtask}~\citep{libtask}, core-aware
Arachne~\citep{arachne}, and language-level threads such as goroutines~\citep{golang} and Java
virtual threads~\citep{loom}). Our fiber runtime is in the second lineage with a fast
register-only switch~\citep{fastwake}, but the contribution is not another threading library: it
is to overlap an \emph{accelerator offload} in servers built either way, quantify the modification
cost across real servers, and map where the transparent form demonstrably stops.

\paragraph{Microsecond-scale systems.}
A body of work rebuilds the OS or runtime to run microsecond-scale tasks efficiently: dataplane
operating systems such as IX~\citep{ix}, work-stealing schedulers such as ZygOS~\citep{zygos}, and
core-reallocating runtimes such as Shenango~\citep{shenango}, Caladan~\citep{caladan}, and
Demikernel~\citep{demikernel}, typically built over kernel-bypass or asynchronous I/O
(DPDK~\citep{dpdk}, \texttt{io\_uring}~\citep{iouring}). They attack the same killer microsecond
problem~\citep{microsecond} but demand a new stack and rewritten applications. We instead ask how
little it costs to overlap one offload inside existing, unmodified servers; the directions are
complementary.

\paragraph{Full hardware offload.}
Another line moves the \emph{entire} datapath onto hardware: KV-Direct~\citep{kvdirect} runs a
key-value store in the NIC, ClickNP~\citep{clicknp} compiles network functions to an FPGA,
iPipe~\citep{ipipe} offloads application logic onto SmartNICs (and must itself schedule the
offloaded tasks' granularity, the dual of our problem), and GPUnet~\citep{gpunet} lets GPU code
drive the network. These eliminate the CPU's role and require rebuilding the application around
the accelerator. Our target is the opposite and far more common case: the application stays on the
CPU and must hide one fine-grained step's latency with almost no change.

\paragraph{Offload integrations and async frameworks.}
Closer to us, specific systems overlap specific offloads---TLS on GPUs (SSLShader~\citep{sslshader}),
QAT engines in nginx's async paths~\citep{qat}, and inference servers that batch GPU work
(RedisAI~\citep{redisai}, Clipper~\citep{clipper})---but each is a point integration tuned to one
server and offload. Async frameworks (libevent~\citep{libevent}, libuv~\citep{libuv},
Seastar~\citep{seastar}, \texttt{async}/\texttt{await}) and proxy offload engines (HAProxy
SPOE~\citep{spoe}, Envoy \texttt{ext\_proc}~\citep{extproc}) make overlap the default, but only if
the application is written in them from the start. We instead inject overlap into existing servers
and predict the win and the cost across the landscape.

\paragraph{Conflict detection.}
Dirty tracking via page protection, software distributed shared memory~\citep{treadmarks}, and
transactional memory~\citep{htm} all detect or prevent conflicting concurrent writes. Our detector
borrows the page-protection technique but is lightweight and specialized to the one hazard
rerouting introduces: a read-modify-write split across an offload.

\paragraph{Symbol interposition.}
The fragility of interposing versioned glibc symbols is folklore among practitioners. Our
condition-variable finding (Appendix~\ref{app:obstacles}) documents a concrete, reproducible
instance and its fix.

% =====================================================================
\section{Conclusion}
\label{sec:conclusion}

What should the CPU do while it waits for the accelerator? It should overlap the offload with
other requests---and it does not need a new framework, runtime, or operating system to do so,
because every server that serves concurrent requests already contains the machinery overlap
requires. Rerouting the offload through that machinery takes tens of lines in off-the-shelf
servers across every concurrency model ($1.2$--$5.4\times$ on real hardware), is predictable in
advance from that model and the offload's weight, can occasionally be done
with zero lines from outside the binary ($17.3\times$, within a precisely characterized envelope),
and stays correct with one targeted guard for the run-to-completion atomicity it suspends. The
result is a practical recipe, and a map, for hiding fine-grained accelerator-offload latency in
the servers that run online services today.

\section*{Acknowledgements}
This paper began as a draft written during the author's internship at Microsoft Research in 2017.
Nine years later, with the help of Pine Copilot and Claude Code, the author finally brought it to completion.
The work was produced using Pine Copilot's voice-directed \emph{whisper coding}
workflow~\citep{pineai2026whispercoding}, in which the author specifies, discusses, and reviews
the work by voice while a coding agent---Claude Code with Claude Opus 4.8 and Claude Fable
5---carries out the planning, coding, experiments, and paper writing.
The author thanks BSQL Networking for hosting the NVIDIA RTX PRO 6000 GPU.

\bibliographystyle{plainnat}
\bibliography{reference}

% =====================================================================
\appendix
\section{AES Block-Size Sweep (detailed values)}
\label{app:blocksize}

Table~\ref{tab:blocksize} gives the full per-size measurements behind the offload-weight curve of
Figure~\ref{fig:weight}, all on a real GPU on an \emph{idle} device (no co-tenant compute), driving a
single-event-loop server (Python \texttt{asyncio} with a 32-thread executor) at 50 concurrent
clients. The AES block size is swept over consecutive powers of two from $4$\,KiB (launch-bound,
tens of microseconds) to $8$\,MiB (bandwidth-bound, $\sim$$2.3$\,ms). For each we report the
measured single-op GPU latency (one offload in flight), the synchronous and asynchronous
throughput, and their ratio; \S\ref{sec:eval} interprets the curve.

\begin{table}[h]
  \centering
  \begin{tabular}{rrrrr}
    \toprule
    block size & GPU latency (\textmu s) & sync (req/s) & async (req/s) & speedup \\
    \midrule
    4\,KiB   & 46.4   & 3750 & 4667 & $1.24\times$ \\
    8\,KiB   & 47.0   & 3688 & 4650 & $1.26\times$ \\
    16\,KiB  & 51.4   & 3657 & 4573 & $1.25\times$ \\
    32\,KiB  & 60.3   & 3501 & 4371 & $1.25\times$ \\
    64\,KiB  & 70.2   & 3326 & 4355 & $1.31\times$ \\
    128\,KiB & 92.7   & 2970 & 4218 & $1.42\times$ \\
    256\,KiB & 126.5  & 2706 & 4128 & $1.53\times$ \\
    512\,KiB & 191.8  & 2454 & 4494 & $1.83\times$ \\
    1\,MiB   & 361.6  & 1538 & 3644 & $2.37\times$ \\
    2\,MiB   & 653.6  & 966  & 2858 & $2.96\times$ \\
    4\,MiB   & 1188.1 & 506  & 1937 & $3.83\times$ \\
    8\,MiB   & 2258.6 & 227  & 1228 & $5.41\times$ \\
    \bottomrule
  \end{tabular}
  \caption{Real-GPU AES block-size sweep on a single-event-loop server (idle GPU). Source:
    \texttt{transparent-runtime/apps/aes\_blocksize\_py\_results.csv}.}
  \label{tab:blocksize}
\end{table}

% =====================================================================
\section{Interposition Obstacles for the Transparent Runtime}
\label{app:obstacles}

Running stock binaries under the \texttt{LD\_PRELOAD} fiber runtime of \S\ref{sec:transparent}
surfaced five obstacles below the architectural walls. Each was resolved once and is recorded here
because any threading-interposing tool will meet them.

\begin{figure}[h]
  \centering
  \includegraphics[width=0.9\linewidth]{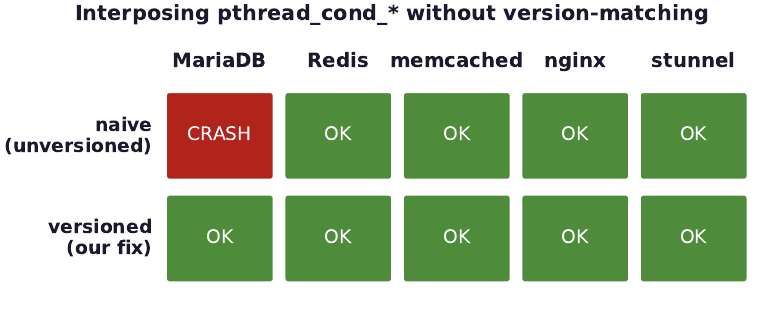}
  \caption{A reusable interposition hazard. A naive unversioned interposer of the glibc
    condition variable silently corrupts \emph{some} binaries (MariaDB crashes at startup) while
    sparing others; version-matching the interposed symbols fixes all of them.}
  \label{fig:condsweep}
\end{figure}

\paragraph{The condition-variable versioning hazard.}
The sharpest obstacle is a \emph{symbol-versioning} hazard~\citep{drepper}: the glibc
condition-variable functions carry two incompatible ABIs under one name
(\texttt{GLIBC\_2.2.5} and \texttt{GLIBC\_2.3.2}), so a naive interposer that defines an
\emph{unversioned} \texttt{pthread\_cond\_signal} breaks the linker's version-matched relocation
and silently corrupts \emph{some} binaries while sparing others: in our sweep it crashes MariaDB
at startup with a wild pointer and is tolerated by Redis, memcached, nginx, and stunnel
(Figure~\ref{fig:condsweep}). Because a transparent runtime cannot know in advance which binaries
are susceptible, version-matched interposition is mandatory: export the interposed symbols at
their exact version via a linker version script and resolve the real ones with \texttt{dlvsym}.

\paragraph{Four lesser obstacles.}
\begin{itemize}[leftmargin=1.4em,itemsep=2pt,topsep=2pt]
  \item \textbf{Alternate I/O entry points.} A server may do its socket I/O through
    \texttt{recv}/\texttt{send}/\texttt{poll} rather than \texttt{read}/\texttt{write}, so every
    blocking entry point the application can reach must be interposed to yield the fiber.
  \item \textbf{Real-thread identity.} \texttt{pthread\_self} must return the genuine OS thread
    identity, because the C library's stack-bounds logic relies on it; returning a per-fiber value
    breaks libc internals.
  \item \textbf{Carrier re-establishment after \texttt{fork}.} A daemon that forks drops the carrier
    thread in the child; the carrier must be re-established before any fiber can run.
  \item \textbf{Priority inversion.} Pinning the carrier to a real-time priority can invert priorities
    against a lock holder running at normal priority, stalling the carrier.
\end{itemize}

\end{document}